\documentstyle[prl,aps,epsfig]{revtex}

\begin{document}
\title{Evidence for charged critical behavior in the pyrochlore superconductor RbOs$%
_{2}$O$_{6}$}
\author{T. Schneider$^{\text{1}}$, R. Khasanov$^{\text{1,2,3}}$,and H. Keller$^{%
\text{1}}$}
\address{$^{\text{(1)}}$ Physik-Institut der Universit\"{a}t Z\"{u}rich,\\
Winterthurerstrasse 190,\\
CH-8057 Z\"{u}rich, Switzerland\\
$^{\text{(2)}}$ Laboratory for Neutron Scattering, ETH Z\"{u}rich\\
and PSI Villigen,\\
CH-5232 Villigen PSI, Switzerland\\
$^{\text{(3)}}$ DPMC, Universit\'e de Gen\`eve, 24 Quai\\
Ernest-Ansermet,\\
CH-1211 Gen\`eve 4, Switzerland}
\maketitle

\begin{abstract}
We analyze magnetic penetration depth ($\lambda $) data of the recently
discovered superconducting pyrochlore oxide RbOs$_{2}$O$_{6}$. Our results
strongly suggest that in RbOs$_{2}$O$_{6}$ charged critical fluctuations
dominate the temperature dependence of $\lambda $ near $T_{c}$. This is in
contrast to the mean-field behavior observed in conventional superconductors
and the uncharged critical behavior found in nearly optimally doped cuprate
superconductors. However, this finding agrees with the theoretical
predictions for charged criticality and the charged criticality observed in
underdoped YBa$_{2}$Cu$_{3}$O$_{6.59}$.
\end{abstract}


\bigskip

Transition metal (TM) oxides are of considerable interest because their
properties range from metal-insulator transitions to colossal
magnetoresistance and superconductivity. TM oxide compounds with the
pyrochlore \ structure have long been studied and have found many
applications\cite{subra}, but it is not until recently that
superconductivity was discovered in one such a material, namely Cd$_{2}$Re$%
_{2}$O$_{7}$ at $T_{c}$ $\approx 1$ K\cite{hanawa,sakai,jin}. Subsequently,
superconductivity was also discovered in the pyrochlore oxides KOs$_{2}$O$%
_{6}$ ($T_{c}\approx 9.6$ K)\cite{yonezawa}, RbOs$_{2}$O$_{6}$ with ($%
T_{c}\approx 6.3$ K)\cite{yonezawarb} and CsOs$_{2}$O$_{6} $ with ($%
T_{c}\approx 3.3$ K)\cite{yonezawacs}. Although the $T_{c}$'s are rather low
the discovery of superconductivity in the pyrochlore oxides opens research
in this area to a new class of materials and raises, of course, the question
of the underlying mechanism. Based on a Cd and Re NMR/NQR study Vyaselev
{\em et al.}\cite{vyaselev} concluded that Cd$_{2}$Re$_{2}$O$_{7}$ behaves
as a weak-coupling BCS superconductor with a nearly isotropic gap, in
agreement with specific heat measurements \cite{hiroi2}. On the other hand,
Koda {\em et al.} \cite{koda} interpreted their penetration depth
measurements on KOs$_{2}$O$_{6}$ in terms of unconventional
superconductivity, while the recent specific heat\cite{bruhw},
magnetization, muon-spin-rotation ($\mu $SR) measurements of the magnetic
penetration depth\cite{khasanov,khasanov2}, and the pressure effect
measurements\cite{khasanov} on RbOs$_{2}$O$_{6}$ reveal consistent evidence
for mean-field behavior, except close to $T_{c}$, where thermal fluctuations
are expected to occur. Hence, a careful study of the thermodynamic
properties of these materials close to $T_{c}$ should allow to discriminate
between mean-field, charged and uncharged critical behavior.

In this work, we focus on RbOs$_{2}$O$_{6}$ and analyze the extended
measurements of the temperature dependence of the magnetic penetration depth
$\lambda $\cite{khasanov2}. Our results strongly suggest that RbOs$_{2}$O$%
_{6}$ falls in the universality class of charged superconductors because
charged critical fluctuations are found to dominate the temperature
dependence of $\lambda $ near $T_{c}$. It differs from the mean-field
behavior observed in conventional superconductors and the uncharged critical
behavior found in nearly optimally doped cuprate superconductors\cite
{ffh,tsda,tshkws,book,parks}, but agrees with the theoretical predictions
for charged criticality\cite{herbut,herbut2,olsson,hove,mo} and the charged
critical behavior observed in underdoped YBa$_{2}$Cu$_{3}$O$_{6.59}$\cite
{ts123charg}.

As long as the effective dimensionless charge $\widetilde{e}=\xi /\lambda
=1/\kappa $ \cite{ffh} is small, where $\kappa $ is the Ginzburg-Landau
parameter, the crossover upon approaching $T_{c}$ is initially to the
critical regime of a weakly charged superfluid where the fluctuations of the
order parameter are essentially those of an uncharged superfluid\cite
{ffh,tsda,tshkws,book,parks}. However, superconductors with rather low $%
T_{c} $'s and far away from any quantum critical point are expected to
exhibit mean-field ground state properties. In this case the Ginzburg-Landau
parameter scales with the specific heat coefficient $\gamma $ as $\kappa
\propto \lambda ^{2}\left( 0\right) \sqrt{\gamma }T_{c}$\cite{gennes}. Thus,
the superconducting pyrochlores, with rather low $T_{c}$'s and moderate $%
\gamma $ and $\lambda \left( 0\right) $ values appear to open up a window
onto the charged critical regime.

\begin{figure}[tbp]
\centering
\includegraphics[totalheight=6cm]{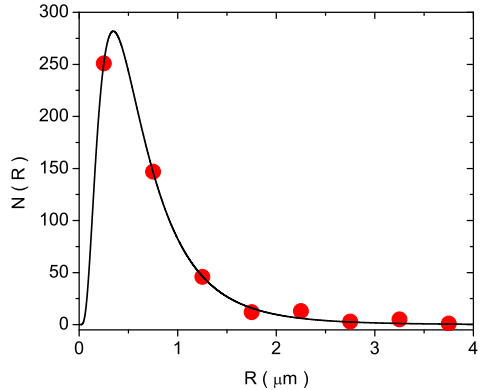}
\caption{Grain size distribution $N\left( R\right) $ of our
RbOs$_{2}$O$_{6}$ sample taken from Khasanov {\em et
al}.\protect\cite{khasanov2}. The solid line is a fit to the
log-normal distribution\protect\cite{kiefer}.} \label{fig1}
\end{figure}

Here we concentrate on the analysis of the magnetic penetration depth data
derived from Meissner fraction measurements\cite{khasanov2}. These data are
in excellent agreement with the $\mu $SR measurements\cite
{khasanov,khasanov2} and sufficiently dense to explore the critical behavior
near $T_{c}$. As required for the determination of the magnetic penetration
depth $\lambda $ from the Meissner fraction measurements the sample was
ground in order to obtain small grains\cite{khasanov2}. Assuming spherical
grains of radius $R$ their size distribution was deduced from SEM (scanning
electron microscope) photographs. The resulting particle size distribution $%
N\left( R\right) $ is shown in Fig.\ref{fig1}.

The data for the temperature dependence of the penetration depth was deduced
by Khasanov {\em et al}. \cite{khasanov2} from the measured Meissner
fraction $f(T)$ by using the Shoenberg formula\cite{shoenberg} modified for
the known grain size distribution\cite{porch},
\begin{equation}
f\left( T\right) =\int_{0}^{\infty }\left( 1-\frac{3\lambda \left( T\right)
}{R}\coth \left( \frac{R}{\lambda \left( T\right) }\right) +\frac{3\lambda
^{3}\left( T\right) }{R^{2}}\right) /\int_{0}^{\infty }g\left( R\right) dR,
\label{eq1}
\end{equation}
where $g\left( R\right) =N\left( R\right) R^{3}$ is the fraction
distribution.

\begin{figure}[tbp]
\centering
\includegraphics[totalheight=6cm]{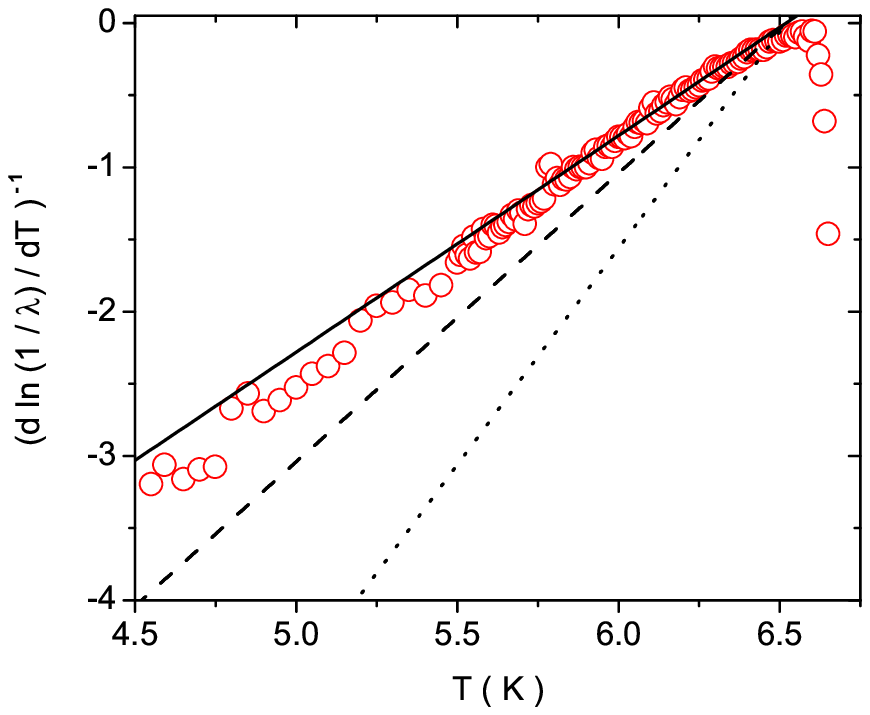}
\caption{$(dln\protect\lambda /dT)^{-1}$ with $\protect\lambda $ in $\protect%
\mu $m versus $T$ for RbOs$_{2}$O$_{6}$ derived from the data of Khasanov
{\em et al}.\protect\cite{khasanov2}. The solid line with slope $1/%
\widetilde{\protect\nu }\simeq 3/2$ corresponds according to Eq. (\ref{eq2})
to charged criticality with $T_{c}=6.52$ K, while the dashed line indicates
mean-field behavior $\left( 1/\widetilde{\protect\nu }\simeq 2\right) $ and
the dotted one 3D-XY critical behavior $\left( 1/\widetilde{\protect\nu }%
\simeq 3\right) $.}
\label{fig2}
\end{figure}

In Fig.\ref{fig2} we depicted the resulting data in terms of $(dln\lambda
/dT)^{-1}$ {\it vs.} $T$. In a homogeneous and infinite system $\lambda $
diverges as
\begin{equation}
\lambda =\lambda _{0}\left| t\right| ^{-\widetilde{\nu }},\text{ }%
t=T/T_{c}-1,  \label{eq2}
\end{equation}
where $\widetilde{\nu }=1/2$ for a conventional mean-field superconductor, $%
\widetilde{\nu }=1/3$ when uncharged thermal fluctuations dominate\cite
{ffh,tsda,tshkws,book} and $\widetilde{\nu }=2/3$ when the charge of the
pairs is relevant\cite{herbut,herbut2,olsson,hove,mo}. Furthermore, when
charged fluctuations dominate the penetration depth and the correlation
length are near $T_{c}$ related by\cite{herbut,herbut2,olsson,hove,mo}
\begin{equation}
\lambda =\kappa \xi ,  \label{eq3}
\end{equation}
contrary to the uncharged case, where $\lambda \propto \sqrt{\xi }$. $\kappa
$ denotes the Ginzburg-Landau parameter. In the plot $(dln\lambda /dT)^{-1}$
{\it vs.} $T$ shown in Fig.\ref{fig2} critical behavior is then uncovered
if\ the data collapse close to $T_{c}$ on a line with slope $1/\widetilde{%
\nu }$. Interestingly enough the data points clearly to charged critical
behavior (solid line with $1/\widetilde{\nu }\simeq 3/2)$, limited by a
finite size effect due to the finite extent of the grains and/or
inhomogeneities within the grains. Indeed, although charged criticality is
attained there is no sharp transition, because $\xi =\lambda /\kappa $
cannot grow beyond the limiting length $L$ and with that $\lambda $ does not
diverge at $T_{c}$.

\begin{figure}[tbp]
\centering
\includegraphics[totalheight=6cm]{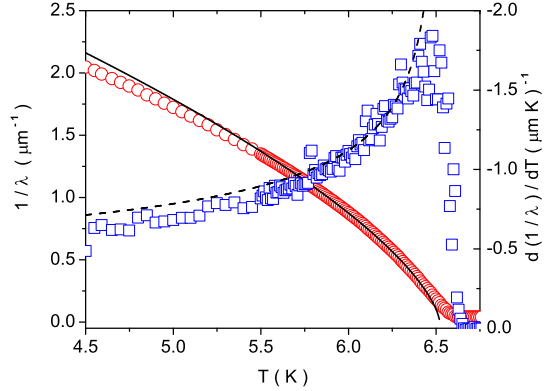}
\caption{$1/\protect\lambda $ and $d(1/\protect\lambda )/dT$ {\it
vs}. $T$ for RbOs$_{2}$O$_{6}$ derived from the data of Khasanov
{\em et al}. \protect\cite{khasanov2}. The solid line is
Eq.\ref{eq2} with $T_{c}=6.52$ K, $\protect\lambda _{0}=0.212$
$\protect\mu $m and $\widetilde{\protect\nu }=2/3$, corresponding
to the charged case, while the dashed line is the derivative.}
\label{fig3}
\end{figure}

To explore the evidence for charged critical behavior and the nature of the
finite size effect further, we displayed in Fig. \ref{fig3} $1/\lambda _{ab}$
and $d(1/\lambda _{ab})/dT$ {\it vs}. $T$. The solid line is Eq. (\ref{eq2})
with $T_{c}=6.52$ K, $\lambda _{0}=0.212$ $\mu $m and $\widetilde{\nu }=2/3$%
, appropriate for charged criticality, and the dashed one its derivative.
Approaching $T_{c}$ of the fictitious homogeneous and infinite system the
data reveals clearly a crossover to charged critical behavior, while the
tail in $1/\lambda $ {\it vs.} $T$ around $T_{c}$ points to a finite size
effect, because $d\lambda /dT$ \ does not diverge at $T_{c}$ but exhibits at
$T_{p}\simeq 6.48$ K an extremum. For this reason $1/\lambda (T)$ possesses
an inflection point at $T_{p}$ and the correlation length attains the
limiting length $L$.

In this case the penetration depth adopts the finite size scaling form \cite
{cardy,schultka}
\begin{equation}
\lambda \left( T\right) =\lambda _{0}\left| t\right| ^{-\widetilde{\nu }%
}g\left( y\right) ,\text{ \ }y=sign\left( t\right) \left| \frac{t}{t_{p}}%
\right| \text{,}  \label{eq4}
\end{equation}
with $\widetilde{\nu }\simeq 2/3$ and $\xi \left( T_{p}\right) =\xi
_{0}\left| t_{p}\right| ^{-\widetilde{\nu }}=L$. For $t$ small and $%
L\rightarrow \infty $ the scaling variable tends to $y\rightarrow \pm \infty
$ where $g\left( y\rightarrow -\infty \right) =1$ and $g\left( y\rightarrow
+\infty \right) =\infty $, while for $t=0$ and $L\neq 0$, $g\left(
y\rightarrow 0\right) =g_{0}\left| y\right| ^{2/3}=g_{0}\left|
t/t_{p}\right| ^{2/3}$. In this limit we obtain
\begin{equation}
\frac{\lambda \left( T_{c},L\right) }{\lambda _{0}}=g_{0}\frac{L}{\xi _{0}}.
\label{eq5}
\end{equation}

\begin{figure}[tbp]
\centering
\includegraphics[totalheight=6cm]{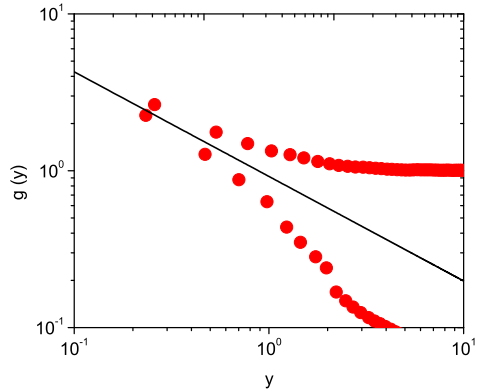}
\caption{Finite size scaling function $g\left( y\right) $ deduced
from the measured data of Khasanov {\em et
al}.\protect\cite{khasanov2} with $T_{c}=6.52$ K, $T_{p}\simeq
6.48$ K, $\protect\lambda _{0}=0.212$ $\protect\mu $m and
$\widetilde{\protect\nu }=2/3$. The solid line indicates the
asymptotic behavior $g\left( y\rightarrow 0\right) =g_{0}\left|
y\right| ^{\protect\nu }$ with $g_{0}=0.92$.} \label{fig4}
\end{figure}

In Fig. \ref{fig4} we displayed the finite size scaling function $g\left(
y\right) $ deduced from the measured data and the parameters emerging from
the fits shown in Fig.\ref{fig3}. \ The solid line indicates the asymptotic
behavior $g\left( y\rightarrow 0\right) =g_{0}\left| y\right| ^{2/3}$ with $%
g_{0}=0.92$. The upper branch corresponding to $T<T_{c}$ tends to $g\left(
\left| y\right| \rightarrow \infty \right) =1$, while the lower one
referring to $T>T_{c}$ approaches $g\left( \left| y\right| \rightarrow
\infty \right) =0$. Consequently, the absence of a sharp transition (see
Figs. \ref{fig2} and \ref{fig3}), is fully consistent with a finite size
effect arising from a limiting length $L$, attributable to the finite extent
of the grains and/or inhomogeneities within the grains. To disentangle these
options we invoke Eq.(\ref{eq5}) yielding with $\lambda \left(
T_{c},L\right) $ $\simeq 6.45$ $\mu $m, $\lambda _{0}\approx 0.212 $ $\mu $m
and $g_{0}=0.92$, $L/\xi _{0}\approx 33$. \ An estimate of $\xi _{0}$ can be
derived from the magnetic field dependence of the maximum of the specific
heat coefficient\cite{tsbled}. As a remnant of the zero-field singularity,
there is for fixed field strength a peak adopting its maximum at $%
T_{p}\left( H\right) $ which is located below $T_{c}$. $T_{p}\left( H\right)
$ is given by $\left( 1-T_{p}\left( H\right) /T_{c}\right) ^{4/3}=\left(
aH\xi _{0}^{2}\right) /\Phi _{0}$, where $a\simeq 3.12$. From the specific
heat data of Br\"{u}hwiler {\em et al.} \cite{bruhw} we derive $\xi
_{0}\approx 0.008$ $\mu $m. Accordingly, $L\approx 33\xi _{0}\approx 0.26$ $%
\mu $m. On the other hand, a glance to Fig.\ref{fig1} shows that the grain
size distribution of the RbOs$_{2}$O$_{6}$ sample exhibits a maximum at $%
\approx 0.69$ $\mu $m\ and decreases steeply for smaller grains. \ Hence,
the smeared transition is not attributable to the finite extent of the
grains but most likely due to inhomogeneities within the grains.

In summary, we have presented an extended analysis of penetration depth data
for RbOs$_{2}$O$_{6}$ near $T_{c}$ providing consistent evidence for charged
critical behavior of the superconductor to normal state transition in type
II superconductors ($\kappa _{0}>1/\sqrt{2}$). The crossover upon
approaching $T_{c}$ is thus to the charged critical regime, while in nearly
optimally doped cuprates it is to the critical regime of a weakly charged
superfluid where the fluctuations of the order parameter are essentially
those of an uncharged superfluid. However, there is the inhomogeneity
induced finite size effect which renders the asymptotic critical regime
difficult to attain \cite{tsrkhk,tsbled,tsdc}. Nevertheless, our analysis of
the penetration depth data of RbOs$_{2}$O$_{6}$ provides remarkable
consistency for critical fluctuations, consistent with the charged
universality class, limited close to $T_{c}$ of the fictitious infinite and
homogeneous counterpart by a finite size effect predominantly due to
inhomogeneities within the grains. Accordingly our analysis strongly
suggests that RbOs$_{2}$O$_{6}$ is not a conventional mean-field
superconductor because charged critical fluctuations dominate the
temperature dependence of the penetration depth near $T_{c}$. These
fluctuations should also affect the specific heat\cite{dasgupta} and the
magnetic properties. However, more extended measurements near criticality
are needed to uncover this behavior.

\acknowledgments The authors are grateful to B. Batlogg, M. Br\"{u}hwiler,
D. Di Castro and A. Sudb\o ~ for useful comments and suggestions on the
subject matter. This work was partially supported by the Swiss National
Science Foundation and the NCCR program {\it Materials with Novel Electronic
Properties} (MaNEP) sponsored by the Swiss National Science Foundation.


\begin{references}
\bibitem{subra}  M. A. Subramanian, G. Aravamudan, and G. V. S. Rao, Prog.
Solid St. Chem. {\bf 15}, 55 (1983).

\bibitem{hanawa}  M. Hanawa, Y. Muraoka, T. Tayama, T. Sakakibara, J.
Yamaura, and Z. Hiroi, Phys. Rev. Lett. {\bf 87}, 187001 (2001).

\bibitem{sakai}  H. Sakai, K. Yoshimura, H. Ohno, H. Kato, S. Kambe, R. E.
Walstedt, T. D. Matsuda, Y. Haga, and Y. Onuki, J. Phys.: Condens. Matter
{\bf 13}, L785 (2001).

\bibitem{jin}  R. Jin, J. He, S. McCall, C. S. Alexander, F. Drymiotis, and
D. Mandrus, Phys. Rev. B {\bf 64}, 180503 (2001).

\bibitem{yonezawa}  S. Yonezawa, Y. Muraoka, Y. Matsushita and Z. Hiroi, J.
Phys.: Condens. Matter {\bf 16}, L9 (2004).

\bibitem{yonezawarb}  S. Yonezawa, Y. Muraoka, Y. Matsushita and Z.Hiori, J.
Phys. Soc. Japan {\bf 73}, 819 (2004).

\bibitem{yonezawacs}  S. Yonezawa, Y. Muraoka and Z. Hiroi, cond-mat/0404220.

\bibitem{vyaselev}  O. Vyaselev, K. kobayashi, K. Arai, J. Yamazaki, K.
Kodama, M. Takigawa, M. Hanawa, and Z. Hiori, J. Phys. Chem. Solids {\bf 63}%
, 1031 (2002).

\bibitem{hiroi2}  Z. Hiroi and M. Hanawa, J. Phys. Chem. Solids {\bf 63},
1021 (2002).

\bibitem{koda}  A. Koda, W. Higemoto, K. Ohishi, S. R. Saha, R. Kadono, S.
Yonezawa, Y. Muraoka, and Z. Hiroi, cond-mat/0402400.

\bibitem{bruhw}  M. Br\"{u}hwiler, S.M. Kazakov, N.D. Zhigadlo, J.
Karpinski, and B. Batlogg, Phys. Rev. B {\bf 70}, 020503 (2004).

\bibitem{khasanov}  R. Khasanov, D. G. Eshchenko, J. Karpinski, S. M.
Kazakov, N. D. Zhigadlo, R. Br\"{u}tsch, D. Gavillet, and H. Keller,
cond-mat 0404542.

\bibitem{khasanov2}  R. Khasanov, D. G. Eshchenko, D. Di Castro, A.
Shengelaya, F. La Mattina, A. Maisuradze, C. Baines, H. Luetkens, J.
Karpinski, S. M. Kazakov, and H. Keller, unpublished.

\bibitem{ffh}  D. S. Fisher, M. P. A. Fisher and D. A. Huse, Phys. Rev. B
{\bf 43}, 130 (1991).

\bibitem{tsda}  T. Schneider and D. Ariosa, Z. Phys. B {\bf 89}, 267 (1992).

\bibitem{tshkws}  T. Schneider and H. Keller, Int. J. Mod. Phys. B {\bf 8},
487 (1993).

\bibitem{book}  T. Schneider and J. M. Singer, {\it Phase Transition
Approach To High Temperature Superconductivity}, (Imperial College Press,
London, 2000).

\bibitem{parks}  T. Schneider, in {\it The Physics of Superconductors},
edited by K. Bennemann and J. B. Ketterson (Springer, Berlin 2004) p. 111.

\bibitem{herbut}  I. F. Herbut and Z. Tesanovic, Phys. Rev. Lett. {\bf 76},
4588 (1996).

\bibitem{herbut2}  I. F. Herbut, J. Phys. A {\bf 30}, 423 (1997).

\bibitem{olsson}  P. Olsson and S. Teitel, Phys. Rev. Lett., {\bf 80}, 1964
(1998).

\bibitem{hove}  J. Hove and A. Sudb\o , Phys. Rev. Lett., {\bf 84}, 3426
(2000).

\bibitem{mo}  S. Mo, J. Hove, A. Sudb\o , Phys. Rev. B {\bf 65}, 104501
(2002).

\bibitem{ts123charg}  T. Schneider, R. Khasanov, K. Conder, E. Pomjakushina,
R. Br\"{u}tsch, and H. Keller, cond-mat/0406691.

\bibitem{gennes}  P. G. de Gennes, {\it Superconductivity of Metals and
Alloys}, (Benjamin, New York 1966).

\bibitem{kiefer}  J. Kiefer and J.B. Wagner, J. Elecktrochem. Soc. {\bf 135}%
, 198 (1988).

\bibitem{shoenberg}  D.~Shoenberg, Proc.~R.~Soc.~Lond. {\bf A 175}, 49
(1940).

\bibitem{porch}  A. Porch {\em et al.}, Phycica C{\bf \ 214}, 350 (1993).

\bibitem{cardy}  J. L. Cardy ed., {\it Finite-Size Scaling}, North Holland,
Amsterdam 1988.

\bibitem{schultka}  N. Schultka and E. Manousakis, Phys. Rev. B {\bf 52},
7528 (1995).

\bibitem{tsrkhk}  T. Schneider, R. Khasanov, K. Conder, and H Keller, J.
Phys. Condens. Matter {\bf 15}, L763 (2003).

\bibitem{tsbled}  T. Schneider, Journal of Superconductivity, {\bf 17}, 41
(2004).

\bibitem{tsdc}  T. Schneider and D. Di Castro, Phys. Rev. B {\bf 69}, 024502
(2004).

\bibitem{dasgupta}  C. Dasgupta and B. I. Halperin, Phys. Rev. Lett. {\bf {47%
}, 1556 (1981). }
\end{references}
\end{document}